\documentclass[twocolumn, 5p]{elsarticle}
\usepackage{tabulary}
\usepackage{amsmath}
\usepackage{multirow}
\usepackage{floatrow}
\usepackage{bigstrut}
\usepackage[hyphens]{url}
\usepackage{units}
\usepackage{eurosym}
\graphicspath{{./figures/}}
\journal{International Journal of Electrical Power \& Energy Systems}

\begin{document}

\begin{frontmatter}
\title{Analysing the Degree of Meshing in Medium Voltage Target Grids - An Automated Technical and Economical Impact Assessment}

\author[Uni,con]{Leon Thurner}
\author[IWES,con]{Alexander Scheidler}
\author[NetzeBW]{Alexander Probst}
\author[Uni,IWES]{Martin Braun}

\address[Uni]{University of Kassel, Germany}
\address[IWES]{Fraunhofer Institute for Wind Energy and Energy System Technology, Kassel, Germany}
\address[NetzeBW]{Netze BW GmbH, Stuttgart, Germany}
\address[con]{contributed equally}

\begin{abstract}
There are different medium voltage (MV) grid concepts with regard to mode of operation and protection system layout. The increasing installation of distributed generation (DG) raises the question if the currently used concepts are still optimal for future power systems. We present a methodology that allows the automated calculation and comparison of target grids within different concepts. Specifically, we consider radial grids, closed ring grids and grids with switching stations. A target grid structure is optimized for each of those grid concepts based on geographical information. To model a realistic planning process, compliance with technical constraints for normal operation, contingency behaviour and reliability figures are ensured in all grid concepts. We present  a multiphase approach to solve the optimization problem based on an iterated local search meta-heuristic. We then economically compare the grids with regards to investment and operational cost for primary and secondary equipment, to analyse which concept leads to the most overall cost-efficient target grids. Since the methodology allows an automated evaluation of a large number of grids, it can be used to draw general conclusions about the cost-efficiency of specific concepts. The methodology is applied to 44 real MV grids spanning about 4800 km of lines, for which the results show that a radial grid structure is overall cost effective compared to grid topologies with switching stations or closed rings even in grid areas with a large DG penetration. The contribution of this paper is threefold: first, a comprehensive methodology to compile automated target grid plans under realistic premises is presented. Second, the practical applicability of the approach is demonstrated by its application in a large scale case study with a high degree of automation. And third, the results of the case study allow to draw conclusions about the techno-economical differences of different MV grid concepts.
\end{abstract}

\end{frontmatter}

\section{Introduction}
The increasing installation of distributed generation (DG) in electric distribution systems leads to a gradual transformation from a centralized to a decentralized power generation. This paradigm shift presents a significant challenge for distribution system operators (DSO), especially in medium voltage (MV) and low voltage (LV) grids. For example, the German DSO Netze BW GmbH expects necessary investment costs between 1.5 and 1.9 billion Euros to integrate additional DG units between the years 2012 and 2030 \cite{EnBW_VNS}. 

\subsection{DG Integration and Increased Degree of Meshing}
Several improvements in grid operation have been considered to reduce grid integration costs, such as controllable MV/LV substations, innovative HV/MV transformer control or intelligent curtailment of DG feed-in \cite{EnBW_VNS}. It has also been suggested that an increased degree of meshing in the MV grid can facilitate the integration of DG by improving the operational behaviour of the grid \cite{Celli2004, Repo2003, Nikander2003}, which could decrease the need for additional grid extension. 
But closed ring systems also require a more sophisticated protection system layout than open ring systems \cite{Viawan2006, Salman2007}, which leads to additional costs. So while meshing can be a technical solution for DG integration, the question arises if it is also economically feasible. It is therefore necessary to study not only the qualitative effect of additional integration of DG by closed ring structures, but to quantify the benefits in terms of saved grid extension and compare them to the additional costs for a more sophisticated protection system layout. Because of the high diversity of distribution grids, it is difficult to draw substantiated conclusions from exemplary case studies. For statistically relevant results, a grid study has to be based on the evaluation of a large number of grids. This can only be achieved with a high degree of automation \cite{Scheidler.2017}.

\subsection{Automated Comparison of Grid Concepts}
In this paper we introduce a methodology that allows the automated comparison of different grid concepts for future power systems. We consider the current grid structure as well as the expected changes in future power systems, such as increasing installation of DG and increasing share of underground cables. We then derive three different possible future grid structures following the three different grid concepts: radial grid without switching station, radial grid with switching station and closed ring grid. To ensure that all grids have equivalent technical capabilities, all grids have to comply with the same technical constraints. These include constraints on voltage and line loading in normal operation and in contingency operation as well constraints that limit the outage times in the grid. The different grid plans are then compared with regard to their cost for primary and secondary equipment. This methodology is applied to 44 real MV grid areas that span about \unit[4800]{km} of lines to reach a sound conclusion about the technical and economic feasibility of each grid concept. The presented approach takes the existing grid infrastructure as well as a prognosis for the future development of the grid into account. The results allow to draw conclusions about strategic decisions in grid planning and adaptation of planning principles.

\subsection{Overview}
The paper is organized as follows: Section \ref{sec:concepts} defines the three grid concepts that are considered in this study and Section \ref{sec:switching_station} gives an overview of the overall methodology that is used to replace switching stations. Section \ref{sec:scenario} introduces the grid data used in the case study and the scenario for future power systems which is applied. Section \ref{sec:methodology} describes the optimization problem and its solution with heuristic random search algorithms in general. Section \ref{sec:optimization} outlines the specific multi-phase target grid optimization approach in detail. Finally, the results of applying the methodology to all grids are presented and interpreted in Section \ref{sec:results}. Section \ref{sec:conclusions} gives a summary and an outlook of how the presented methodology can be used in future studies.

\section{Grid Concepts} \label{sec:concepts}
\begin{figure}[tbp]
    \centering
    \includegraphics[width=1.\linewidth]{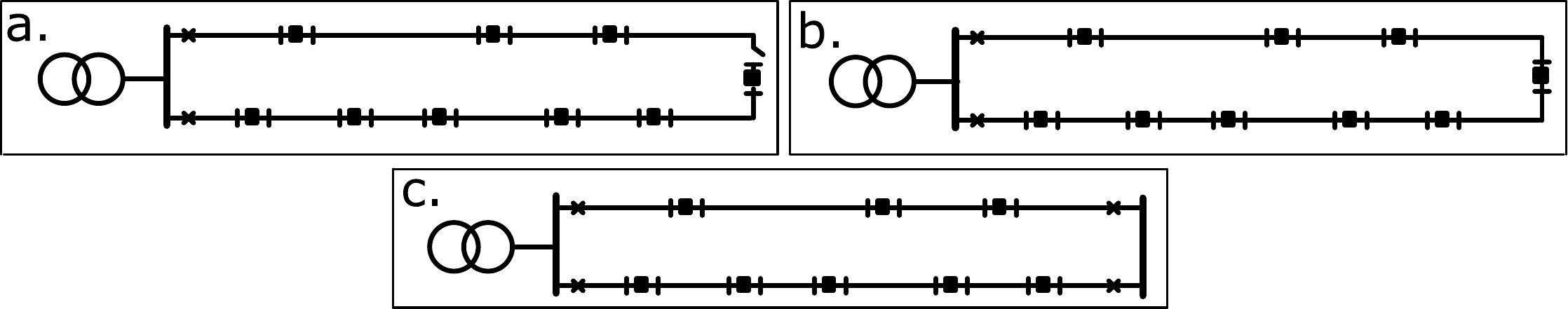}
    \caption{Basic structural MV grid elements: a. open ring, b. closed ring, c. switching station ring.}
    \label{meshing}
\end{figure}

Grid layout and protection system layout have to be inter-coordinated to guarantee a safe grid operation \cite{Schlabbach.2008}. For the grid layout, we consider three basic  structural elements: open line rings, closed line rings and ring structures with switching stations (see Figure \ref{meshing}). A grid concept is then defined as the combination of these elements with an appropriate protection system layout.

Every DSO defines its grid concept depending on individual requirements and boundary conditions. For this study, we consider three different MV grid concepts:

\paragraph*{Radial grid} A radial grid is built exclusively out of open ring structures as shown in Figure \ref{meshing} a. The radial structure ensures that short-circuit currents are always uni-directional, which makes fault detection and location easier than in grid structures with closed rings. We therefore consider a simple protection system layout with over-current protection systems at HV/MV substations (primary substations) for fault cut-off and non-directed short-circuit indicators at the MV/LV substations (secondary substations) for fault location in this concept. 

\paragraph*{Closed Ring grid} A closed ring grid has the same structure as a radial grid, only that some of the rings are operated as closed rings as shown in Figure \ref{meshing} b. Since this allows bi-directional short-circuit currents, the protection system needs to be more sophisticated than in the open ring case. We therefore assume an impedance protection system in the primary substations and directional short-circuit indicators in the secondary substations.

\paragraph*{Switching Station grid} The switching station concept is a radial grid concept with additional MV substations placed in load or generation centres to stabilize the grid. Switching stations are MV substations with over-current protection devices that allow parallel operation of cables (see Figure \ref{meshing} c.). Apart from the additional protection system in the MV substations, the protection system layout is the same as in a radial grid. Using switching stations can be seen as a way to allow closed rings in the grid without changing the overall protection system layout by installing additional circuit breakers in the MV grid.

\begin{figure*}[t]
\includegraphics[width=1.\linewidth]{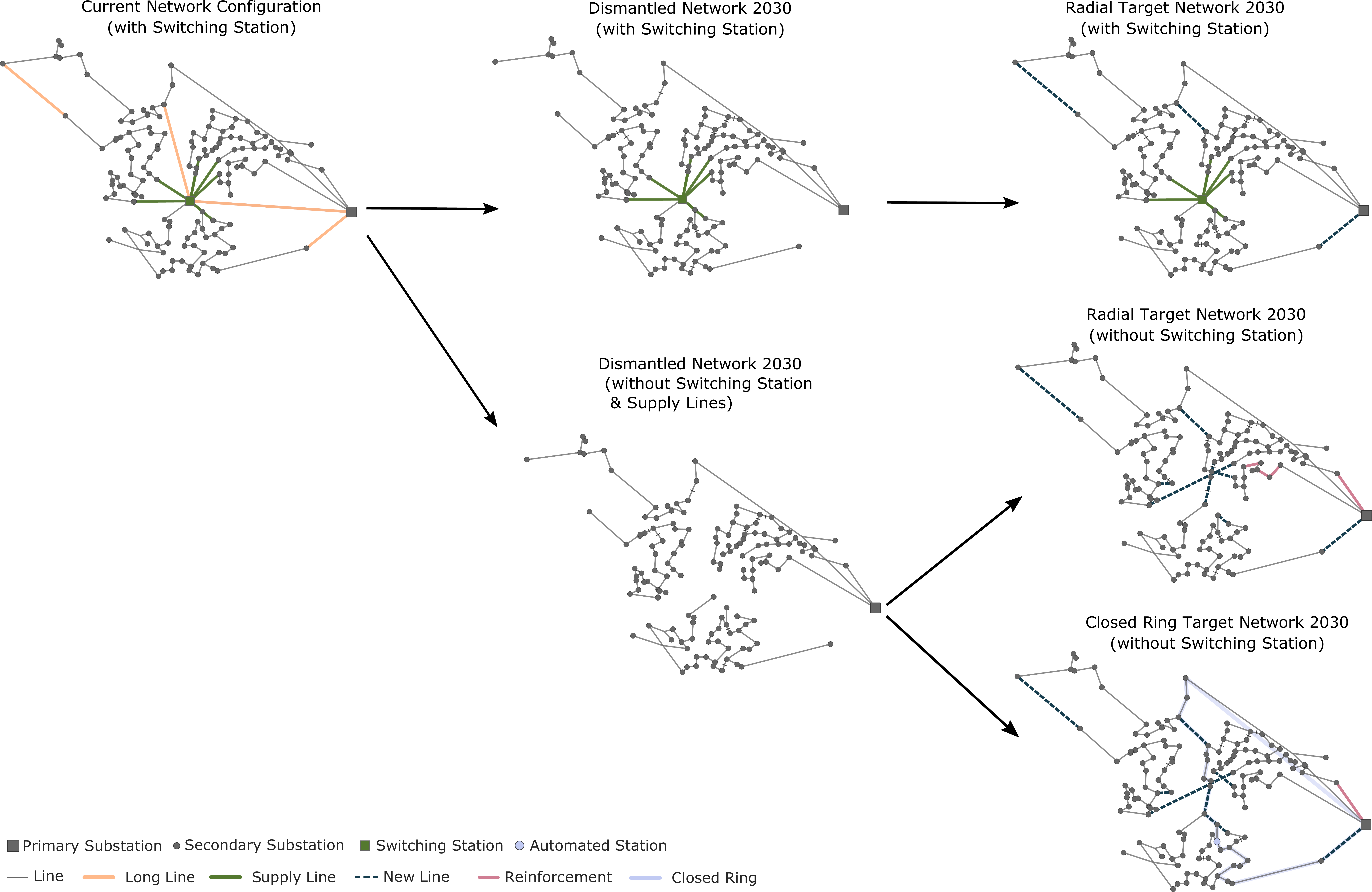}
\caption{Overview of target grid optimization methodology: the current grid structure (left) is dismantled with and without switching station with assumptions for the grid state in 2030 (middle) and then target grids for switching station, open ring and closed ring grid concept is optimized (right)\label{concept_example}}
\end{figure*}

\section{Methodology for Replacing Switching Stations} \label{sec:switching_station}
We use the existing switching stations as starting point for the analysis and optimize the respective area for the three concepts. We assume that the switching station will have to be renewed within the planning horizon, so that a replacement investment will be necessary if the switching station topology is maintained. The question then arises, if it is more economical to avoid this replacement investment by changing the grid concept to a strictly radial or a closed ring grid structure. Founding new switching stations in grid areas that are currently radial is out of the scope of this paper.
An example for the methodology can be seen in  Figure \ref{concept_example}, where a current grid structure with switching station is shown on the left. The grid area around the switching station is now dismantled with respect to the planning horizon of the year 2030 (Figure \ref{concept_example}, middle). Since asset data of the lines was not available, the dismantling is based on the importance of the line for the grid topology. Short lines which connect two secondary substations will be necessary in any grid to ensure the supply of all stations. Longer supply lines however might be replaced by different cable routes depending on the chosen concept. We therefore remove all long lines, which are defined as lines with a length of over \unit[2]{km} for this case study. The methodology works the same with other criteria for which lines should be removed, for example by installation date or standard type if this data is available. 
To develop an alternative radial grid structure, we create a second dismantled grid where the switching station with all of its supply lines is removed in addition (Figure \ref{concept_example}, middle). From these two topologies, we then generate target grids for open ring, closed ring and switching stations concepts as shown in Figure \ref{concept_example} on the right. The optimization process that is used to create these grids is outlined in detail in Section \ref{sec:optimization}.

\section{Grid Data and Scenario} \label{sec:scenario}
In this paper we study 5 MV grid groups at the \unit[20]{kV} voltage level which are connected to the \unit[110]{kV} voltage level through 51 HV/MV transformers. The grids cover a total of about \unit[4800]{km} of lines and service loads with about \unit[770]{MVA} and DG with about \unit[750]{MVA} installed power (see Table \ref{tab:ng}). The grids are all located in southern Germany and operated by the DSO Netze BW GmbH. All MV grids are currently operated as an open ring grid with switching stations in load or generation centres. There are a total of 49 switching stations in the grid data. The grid area around each switching station is taken as a starting point for the analysis as outlined in Section \ref{sec:switching_station}. We model the situation for the planning horizon of 2030 by applying a load prognosis as well as installing additional DG based on a prognosis for the expected \begin{table}[h]
\caption{Parameters of considered grid groups}
\label{tab:ng}
\noindent 
\renewcommand{\arraystretch}{1.0}
\begin{tabulary}{\linewidth}{c c c c c c}
\shortstack{\vspace{0.2em}Grid \\ Group}  &  \shortstack{\vspace{0.1em}HV/MV \\ Transf.} & \shortstack{Switching \\ Stations} & \shortstack{\vspace{0.1em}Load \\ $[$MVA$]$} & \shortstack{\vspace{0.1em}DG \\ $[$MVA$]$} & \shortstack{\vspace{0.1em}Lines \\ $[$km$]$}\\ \hline
 1 & 15 & 16 &  249 & 183 & 1363\\
 2 & 8 & 8 & 87 & 275 & 907 \\
 3 & 8 & 8 & 151 & 231 & 1329 \\
 4 & 12 & 7 & 169 & 28 & 655\\
 5 & 8 & 10 & 117 & 34 & 525\\ \hline
\textbf{Total}& \textbf{51} & \textbf{49} & \textbf{773} & \textbf{751} &  \textbf{4779}

\end{tabulary}
\end{table} expansion of photovoltaic systems (PV) and wind power plants \cite{EnBW_VNS}. Table \ref{tab:ng} shows that the load remains predominant in some grids while in others DG clearly dominates. The set of grid groups was chosen in order to form a good representation of the expected situation in 2030. The DSOs internal planning principles dictate that new line trails are always built as underground cables. We therefore assume that all new lines are built as cables and overhead lines will be replaced by underground cables until the planning horizon in 2030. The grid is modelled and analysed in the open source power system analysis tool pandapower \cite{pandapower}.

\section{Heuristic Grid Optimization} \label{sec:methodology}
The goal of this paper is to economically compare different grid concepts taking into account the currently existing grid structure. Figure \ref{concept_example} shows an example of such a topology optimization. This section outlines, how the optimization problem is formulated and solved with a heuristic random search algorithm.

\subsection{Technical Constraints}
An economic comparison is only valid if the compared grids are  equivalent with respect to their technical capabilities. To ensure all grids meet the technical requirements, we introduce constraints regarding grid topology, operational behaviour in normal operation, operational behaviour in contingency operation and outage times. The goal of the optimization is then to find a grid that complies with all these constraints at the minimal possible costs. Which constraints are applied in the different optimization phases is explained in detail in Section \ref{sec:optimization}.

\subsection{Measures}
To generate realistic grid structures, all options that a grid planner has in the planning process have to be considered as degrees of freedom in the optimization. In this study, we consider the following measures:
\begin{itemize}
\item Replacing existing overhead lines with cables 
\item Replacing existing cables with cables of higher diameter
\item Optimizing the position of normally open switches (sectioning points)
\item Adding cables in parallel to existing line trails
\item Adding cables in new line trails
\item Equipping secondary substations with communication links to improve reliability
\item Removing or renewing switching stations
\end{itemize}
Each of the possible measures can either be applied or not applied, which leads to the formulation of the Distribution system Expansion Problem (DEP) as a combinatorial optimization problem \cite{Scheidler.2017}. The details of the different measure and at what point in the optimization process they are deployed is explained in detail in Section \ref{sec:optimization}.
\begin{table}[t!]
\caption{Annual costs of elements considered in this study}
\label{tab:costs}
\noindent 
\begin{tabulary}{\linewidth}{ll}
Grid Component & Annual Cost \\ \hline
Cable NA2XS2Y 3x1x300 & 7,000 \euro/km\\
MV switching station & 35,900 \euro\\
Communication link & 1,200 \euro \\
Directed short-circuit indicator (update) & 30 \euro/Station \\
Impedance protection (update) & 400 \euro/Feeder \\
\end{tabulary}
\end{table}

\subsection{Cost Assumptions}
Different measures, such as cables or communication links, differ in operational costs, investment costs and life expectancy. To make the costs of these measures comparable, we chose the Net Equivalent Uniform Annual Cost \cite{Hossein2011} as a measure of cost for all grid elements. The annual costs $C_{pa}$ are equal to the sum of annual operational costs $\unit[C_{op, pa}]{[\nicefrac{EUR}{a}]}$ and annual investment costs $C_{inv, pa} [\nicefrac{EUR}{a}]$:
\begin{equation}
C_{pa} =  C_{op, pa} + C_{inv, pa}
\end{equation}
While operational costs are usually already given as yearly costs, investment costs are typically given as total costs $C_{inv, total} [EUR] $. The annual investment costs can be calculated by discounting the total investment costs with the calculation interest rate $\unit[i]{[\%]}$ over the life expectancy $\unit[m]{[a]}$ of the component:
\begin{equation}
C_{inv, pa} = C_{inv, total} \cdot \frac{(1+\nicefrac{i}{100})^{m} \cdot \nicefrac{i}{100}}{(1+\nicefrac{i}{100})^{m} -1}
\label{neuac}
\end{equation}
The cost assumptions used for all relevant components in this study are depicted in Table \ref{tab:costs}. The cable costs are assumed to be constant per kilometre for both adding new lines an replacing existing lines. This is because the main cost factor is the excavator work, which has to be carried out when replacing an existing line as well as when laying a new cable. The full switching station costs are assumed even if the switching station already exists in the current grid, in accordance with the strategic viewpoint of expected renewal outlined in Section \ref{sec:switching_station}.

\subsection{Heuristic Solution}
We formulate the DEP as a combinatorial optimization problem with non-linear constraints. Given a fixed number of possible measures, the goal is to find the cheapest subset of these measures that fulfils all constraints. It is not considered feasible to formulate this optimization problem in a closed form and solve it analytically. Instead, heuristic algorithms such as genetic algorithms \cite{Camargo2013, Miranda1994, Falaghi2011}, evolutionary algorithms \cite{Zmijarevic.2005}, tabu search \cite{Cossi2010, Benvindo2013}, particle swarm optimization \cite{Ganguly2010, Sahoo2012} or artificial immune systems \cite{Keko.2004} are popular methods for its solution. Most of these studies assume the radial structure of the grid as a constraint \cite{Boulaxis2002, Camargo2013, Cossi2010, Diaz2002, Benvindo2013, Falaghi2011, Miranda1994, Sahoo2012} and are therefore not designed to optimize and compare grids in other grid concepts. The approach described in \cite{Scheidler.2017} is however flexible enough to handle the heuristic optimization with varying measures and constraints, so that a study of different grid concepts is possible. The DEP is solved using an Iterated Local Search (ILS) algorithm with Hill Climbing as local search. This paper concentrates on the application of the optimization framework to the specific questions of structural optimization in different grid concepts. Detailed information on the formulation of the optimization problem, codification of measures, definition of the step-wise cost function and neighbourhood function can be found in \cite{Scheidler.2017} and  \cite{Thurner.2017}.
 
\section{Multiphase Target Grid Optimization}  \label{sec:optimization}
\begin{figure}[t]
    \centering
		\includegraphics[width=1.\linewidth]{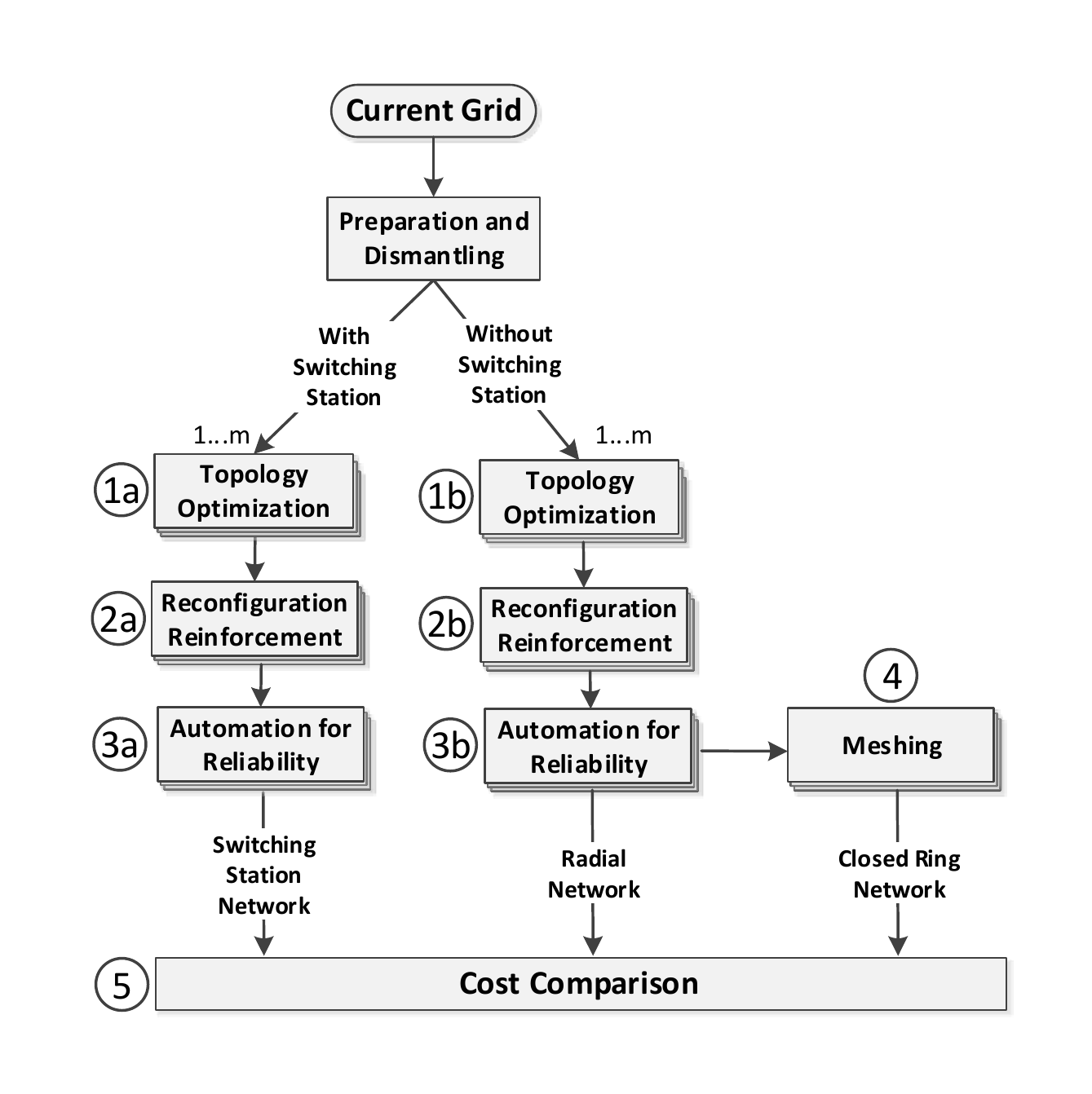}
    \caption{Methodology for the calculation and comparison of the three grid structures}
    \label{gesamt}
\end{figure}

Because of the complexity of the combinatorial optimization problem, it is not feasible to optimize sectioning points, new line trails, line replacement, parallel lines and secondary substation automation all in one optimization. We therefore split the problem into several sub-problems that are solved subsequently. An overview of the multi-phase optimization process that is used to find grid structures for the three concepts starting from the current grid configuration is shown in Figure \ref{gesamt}. Each phase is explained in this section with the help of the example grid area shown in Figure \ref{concept_example}.

\subsection{Phase 1: Topology Optimization}  \label{sec:topology}
A valid grid has to comply with several constraints regarding its structural layout. In the first step, we therefore optimize the topology of the grid.
\begin{figure}[t]
    \centering
		\includegraphics[width=1.\linewidth]{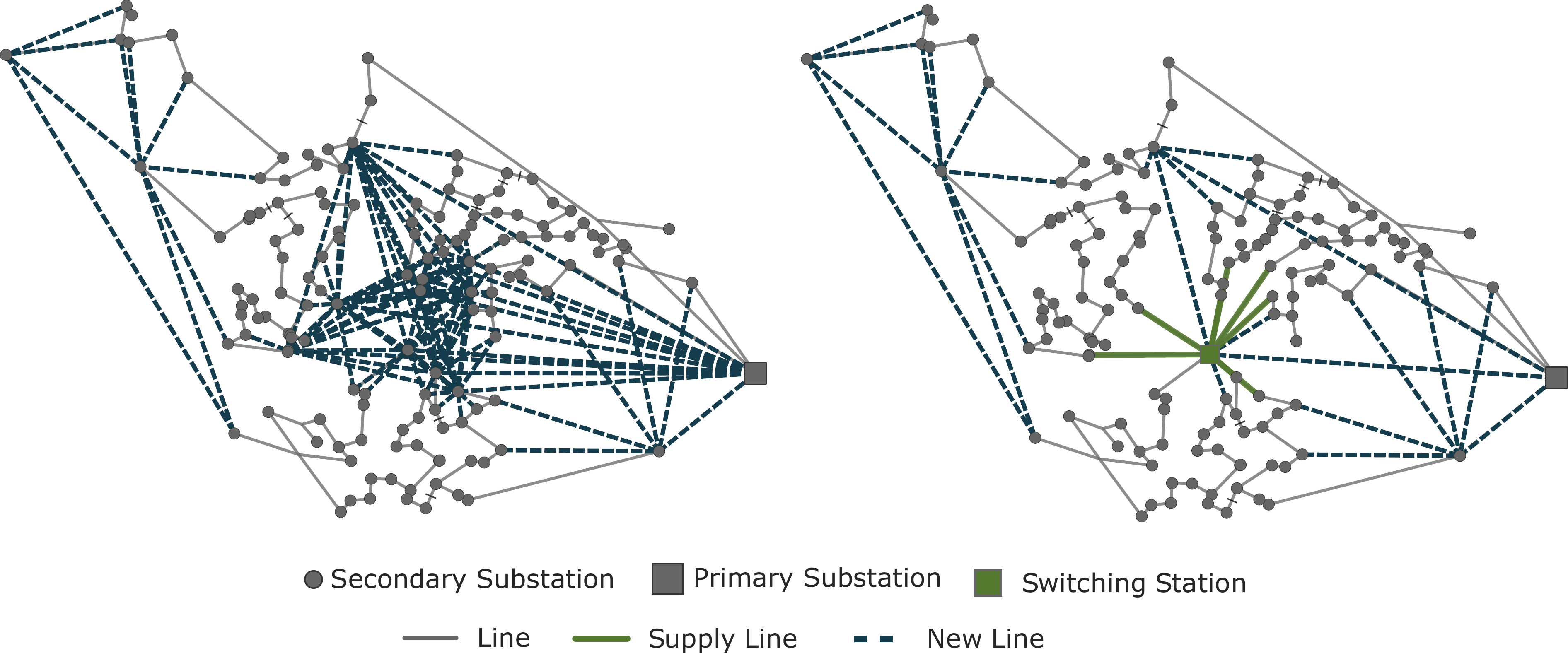}
    \caption{Considered new line trails in the example grid area}
    \label{gis}
\end{figure}
\subsubsection{Topological Constraints} \label{sec:topology_constraints}
Topological constraints specify requirements for the topological structure of the grid without taking electric parameters into account. They can be analysed with graph searches using the pandapower topology package \cite{pandapower}. Which topological constraints are relevant depend on the grid concept and protection system layout. Here, we assume the following constraints:

\paragraph*{Supply connection} All secondary substations have to be connected to an HV/MV substation to be supplied. This means there has to be at least one path from each station to an external grid connection.

\paragraph*{Contingency supply} All secondary substations must have a second connection for resupply in case of a single contingency, which is provided by ring structures in the grid. Secondary substations which are located on stubs and therefore do not have a contingency connection in the current configuration are exempted from this rule. Note that this constraint only guarantees the topological possibility of a resupply, operational constraints in the contingency state are addressed in section \ref{sec:n1_operation}.

\paragraph*{Radiality} The grid has to be radial, in that no feeders can be galvanically connected through closed rings in the MV grid. Rings are only permitted if they go through a switching station. Closed ring structures as shown in Figure \ref{meshing} b. are note permitted, as they will be investigated separately in phase 4 (see Section \ref{sec:phase4}).

\subsubsection{Line Trail Optimization}
To construct grid structures that comply with the constraints outlined above from a dismantled grid, new lines are added between two secondary substations which are not currently connected. The length of a new line is calculated as the air-line distance between the stations multiplied with a factor of 1.5. The costs per kilometre line are given in Table \ref{tab:costs}. We find possible new line trails to complete the grid with a Delaunay-triangulation \cite{deBerg2008} between all stations from which a line was removed. The considered line trails in the example grid are shown in the example in Figure \ref{gis} for the layout with and without switching station. Since there are a lot of possible combinations and the topologies cannot be easily compared, we consider several different solutions. Specifically, we create 50 different topologies for each grid area. Figure \ref{reinforced} shows three of the 50 topologies for the example grid area. All steps that are explained in the following are conducted 50 times per studied grid area.

\subsection{Phase 2: Reconfiguration and Reinforcement} \label{sec:operation}
After the first optimization step, the grids comply with topological constraints, so that all stations are supplied and the grid is radial. In the second optimization step, it is ensured that the electric parameters of the grids comply with the grid planning principles. This is achieved by grid reconfiguration and reinforcement.

\subsubsection{Normal Operation Constraints} \label{sec:normal_operation}
Constraints for equipment loading and bus voltages have to be complied with in every possible load or generation scenario. We therefore define two worst-case scenarios for the peak load and peak generation situation. The simultaneous factors used to generate the worst case scenarios can be seen in Table \ref{tab:simultan}. The scaling factors for the loads refer to the drag pointer measurements at the secondary substations and to the installed power of DG. Since we assume V(P) transformer control, the voltage set point for the transformer voltage controller depends on the active power flow and is therefore different in the two worst-case scenarios. Adherence with the power flow constraints in normal operation is checked with one power flow for each worst-case scenario using pandapower \cite{pandapower}. In both cases, all bus voltages have to be within the voltage band defined by the DSO and the line loadings have to be below \unit[100]{\%}.
\begin{table}[t]
\noindent 
\renewcommand{\arraystretch}{1.0}
\begin{tabulary}{\linewidth}{lll}
  &  Peak Load  & Peak Generation  \\ \hline
Scaling Load & 1.0 & 0.3 \\
Scaling DG (PV) & 0.0 & 0.8 \\
Scaling DG (Wind) & 0.0 & 1.0 \\
Transformer Setpoint & \unit[1.0]{pu} & \unit[1.05]{pu}\\
\end{tabulary}
\caption{Definition of worst-case scenarios \label{tab:simultan}}
\end{table}
\subsubsection{Constraints in Contingency Operation} \label{sec:n1_operation}
The Single Contingency Policy (SCP) dictates, that the power system must continue functioning after the outage of one power system element. The topological contingency constraint ensures that a resupply is possible after the failure of a line through a backup connection (see Section \ref{sec:topology_constraints}). However, it also has to be ensured that the grid operates safely in the resupplied state. Specifically, the bus voltages have to be within the contingency voltage band (which is usually broader than the voltage band for normal operation) and the line loadings have still to be below \unit[100]{\%}.

\begin{figure}[t]
    \centering
    \includegraphics[width=1.\linewidth]{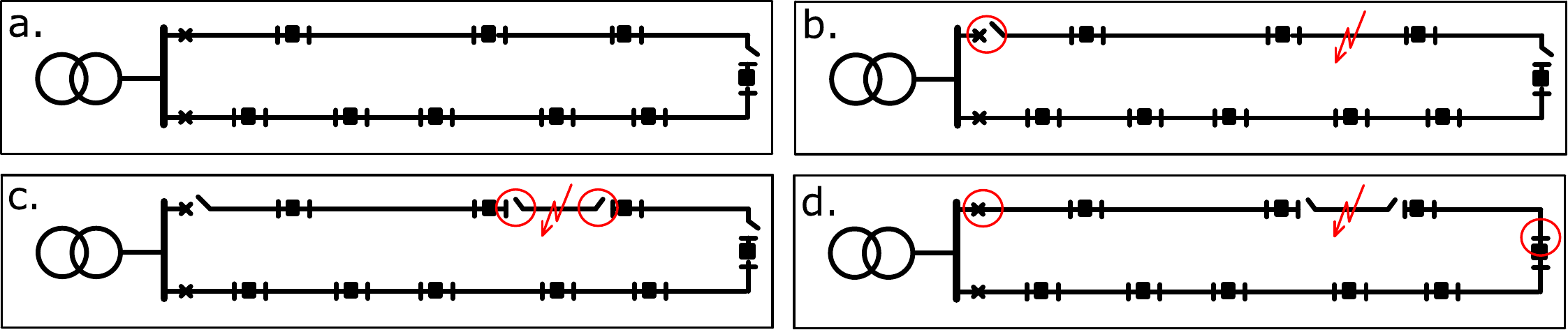}
    \caption{Resupply switching sequence for a line fault in an open ring: a. normal operation, b. line fault with protection trip, c. fault isolation, d. resupply}
    \label{resupply}
\end{figure}

To check this constraint, we calculate a resupply scenario for each feeder line directly connected to transformer or switching station. A fault of such a line is the worst case scenario, since the whole ring has to be supplied through only one feeder. A typical switching sequence for a line outage in a radial grid is shown in Figure \ref{resupply}. When a line fails, the circuit breaker in the primary substation opens to disrupt the short-circuit current, and thereby cuts the affected half ring from power supply (Figure \ref{resupply} b.). After the fault is located, it is isolated by manually opening the load-break switches at the secondary substations connected to the faulted line segment (Figure \ref{resupply} c.). Finally, the circuit breaker in the primary substation as well as the sectioning point are closed to allow a resupply of all stations on the ring (Figure \ref{resupply} d.). This switching sequence is automatically carried out for all feeders using the pandapower topology package to find a resupplied state. Compliance with contingency constraints is checked with a power flow in the resupplied state. The constraints are checked for the peak load case as defined in Table \ref{tab:simultan}. The peak generation scenario is not checked in contingency situations because DG are not considered to be subject to the SCP.

\subsubsection{Optimization of Operational Behaviour}
We consider switching and cable measures to improve the electric parameters in a grid structure and ensure compliance with constraints for normal and contingency operation.

\paragraph*{Sectioning Point Optimization} In a radial grid structure, sectioning points are necessary to fulfill the topological constraints, specifically to avoid direct galvanic connection of feeders or transformers. Sectioning points can be relocated to improve the operational behaviour of the grid. The heuristic optimization aims to find a switching state which minimizes the voltage and line loading violations while still complying with all topological constraints. Details about the used composite objective function are described in \cite{Thurner.2017}.

\paragraph*{Cable Measures} Existing line segments can be replaced by a new cable with higher ampacity, to mitigate line overloading, or with a lower impedance, to mitigate voltage problems. If line replacement is not enough to mitigate all constraint violations, additional lines can be added to the grid in parallel to existing line routes. Parallel lines can only be connected to secondary substations and the radiality constraints have to be respected.

Constraint violations are first mitigated through sectioning point optimization, since opening and closing switches is not associated with costs. If not all constraint violations can be mitigated through reconfiguration, the grid has to be reinforced with additional cable measures until all violations are mitigated.

\begin{figure}[t]
    \centering
		\includegraphics[width=.47\linewidth]{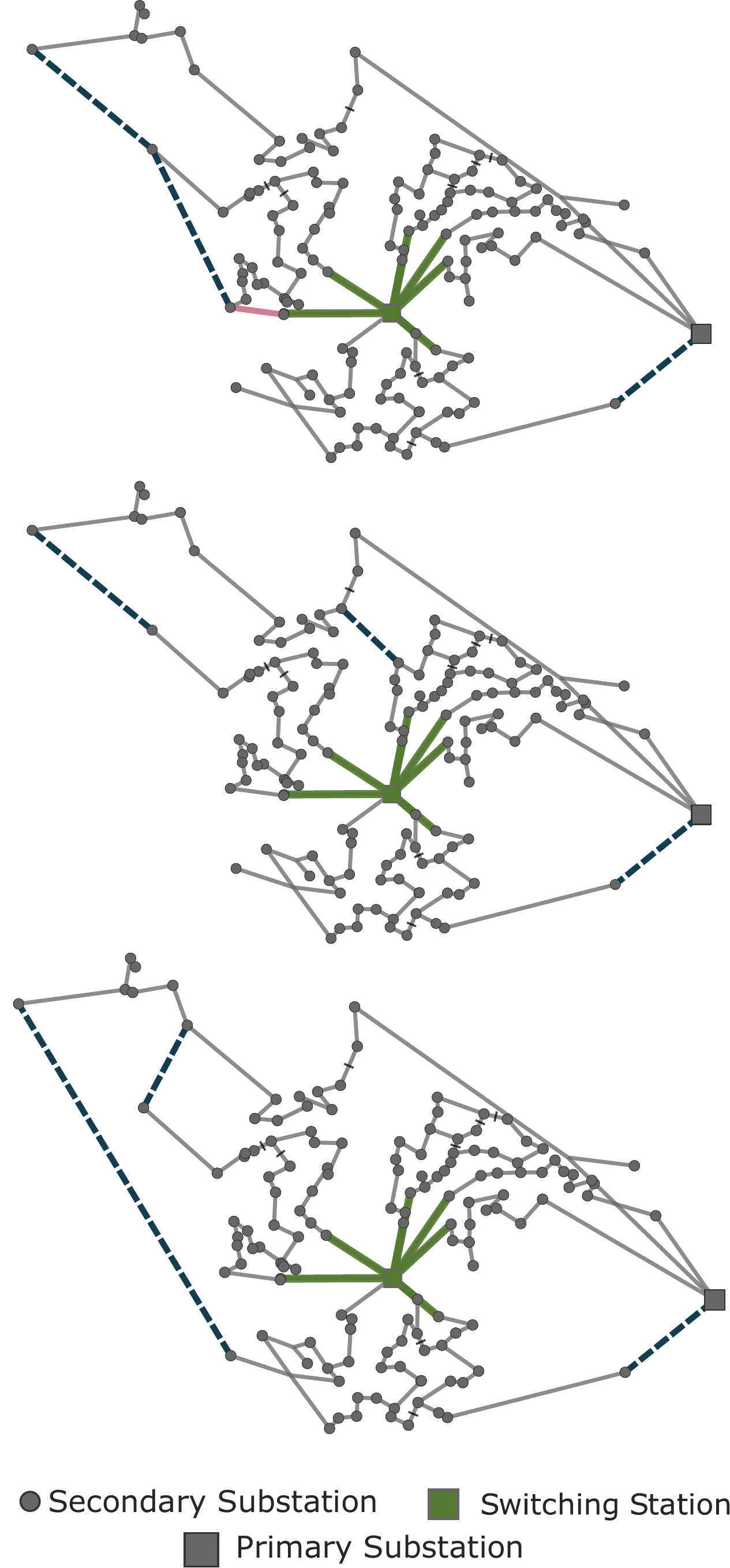}
    \includegraphics[width=.47\linewidth]{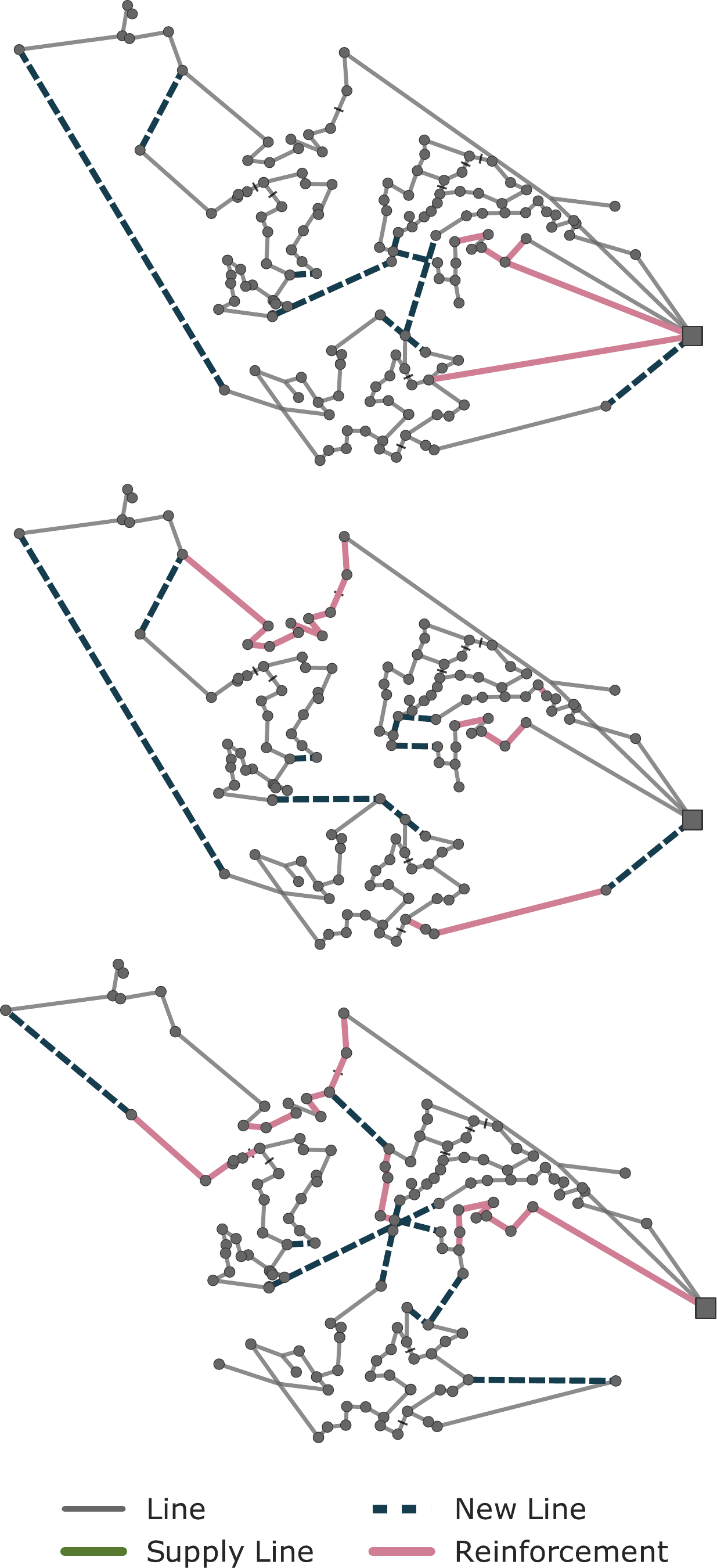}
    \caption{Necessary reinforcement for the three example grids with and without switching station}
    \label{reinforced}
\end{figure}

The necessary reinforcement for the three example grids can be seen in Figure \ref{reinforced}. It can be seen that there is overall less need for reinforcement in the switching station grid, since the switching station serves as a junction that stabilizes the grid.

\subsection{Phase 3: Automation for Reliability}
After the second optimization step, all grids comply with topological constraints as well as operational constraints for normal and contingency operation. To ensure that the grids are also sufficiently reliable, the grid is equipped with communication links to speed up resupply and reduce outage times if necessary.

\begin{figure*}[tbp]
    \centering
		\includegraphics[width=.465\linewidth]{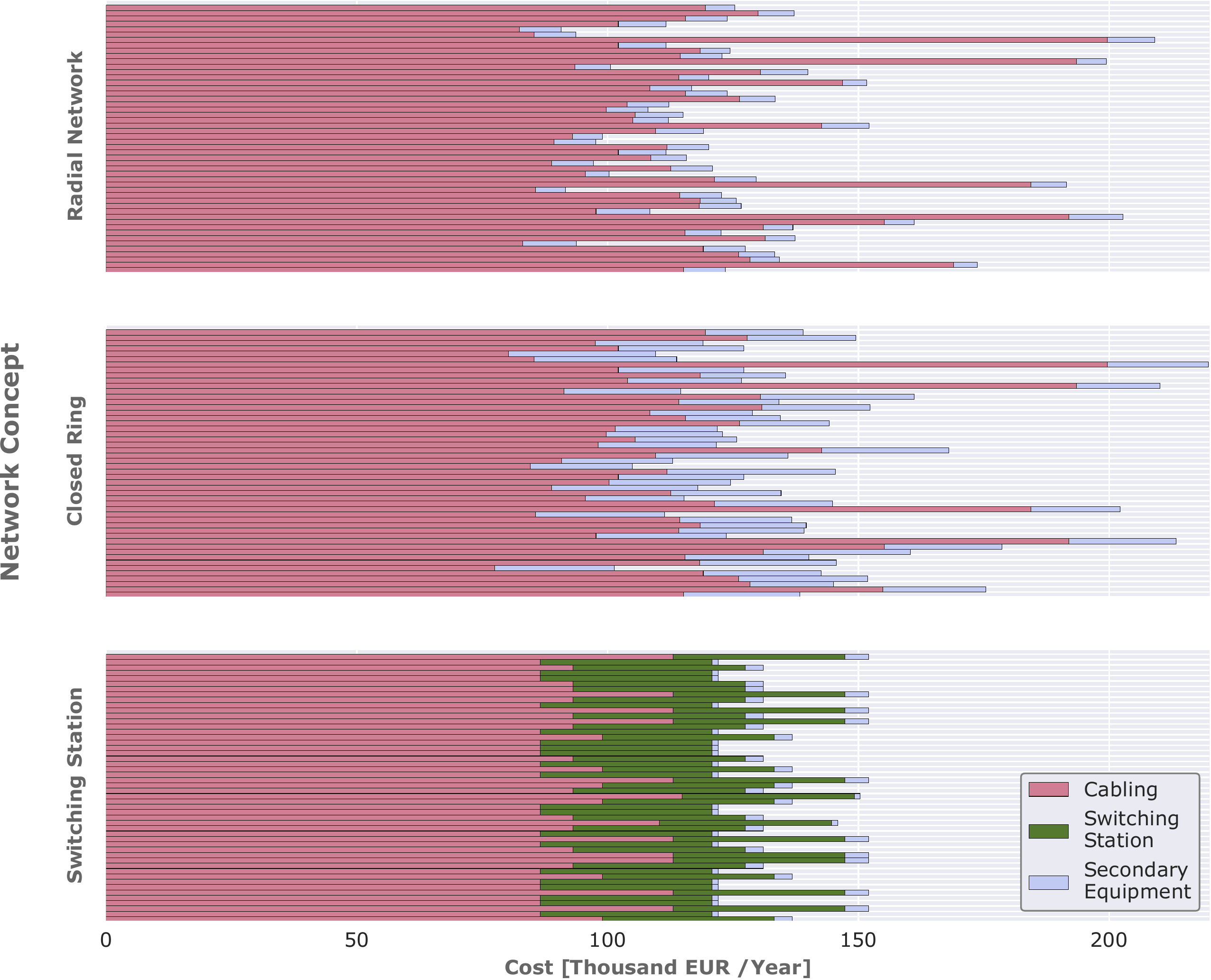} \hspace{0.1em}
		\includegraphics[width=.175\linewidth]{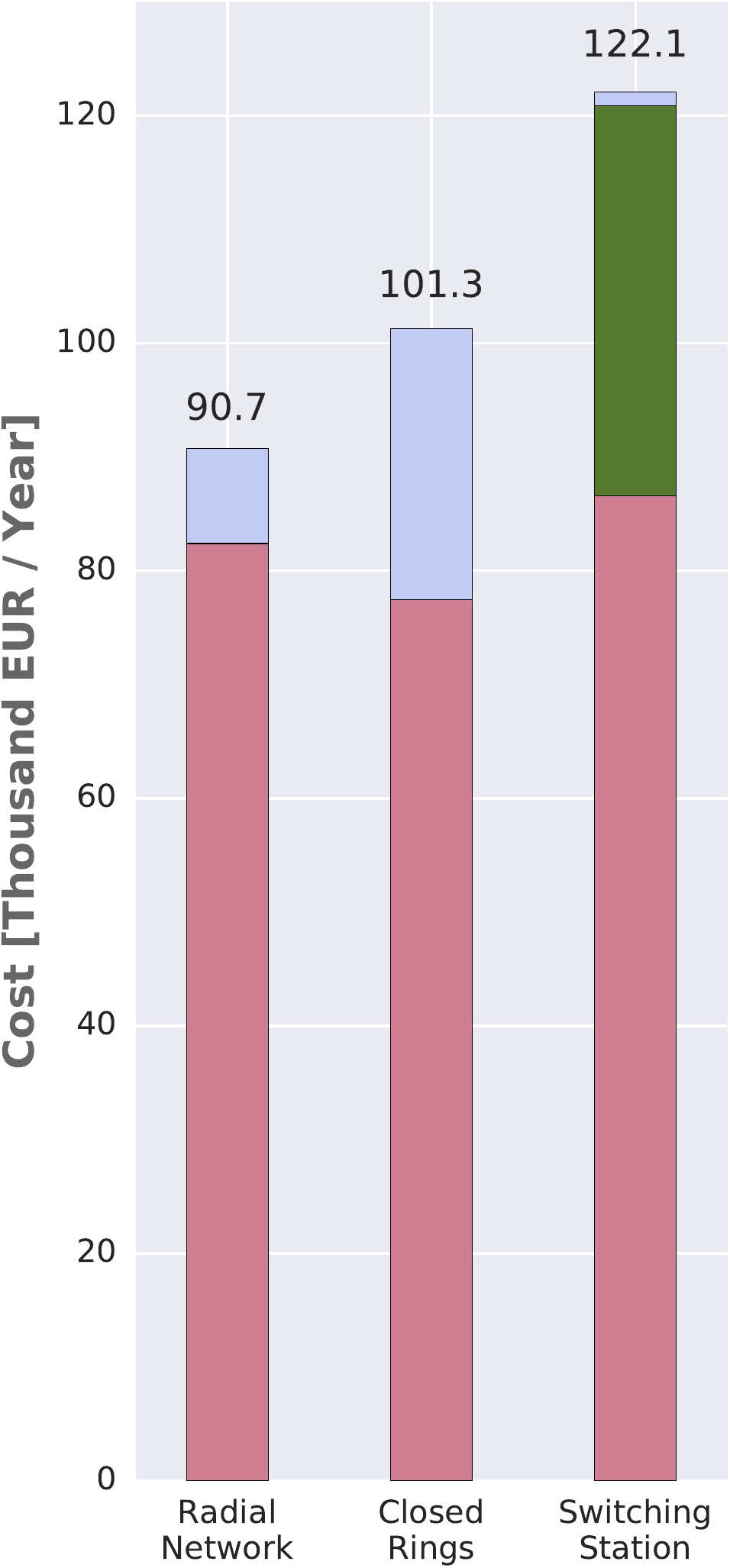}

    \caption{Cost comparison for example grid area: costs for all 50 separately optimized grids (left) and cost minimal reference grid (right) for each concept}
    \label{cost_comparison}
\end{figure*}

\subsubsection{Reliability Constraints}
The SCP demands that the grid complies with the operational constraints in the resupplied state. It does however not draw any conclusion about the frequency of fault occurrences in the grid or about which secondary substations will experience outages for how long. In this optimization step, it is ensured that the grid structure is sufficiently reliable with respect to outage frequency and restoration times. The service reliability can be measured in reliability figures, the most prominent of which is the Expected Average System Interruption Duration Index (ASIDI), which can be computed with a Failure mode and effects analysis (FMEA) \cite{Brown2009}. The FMEA takes into account the failure rates of components as well as the effect of the failure. The failure rates of the components are taken from statistical analysis where it is given as a function of the line type and, in case of underground cables, the insulation material. The effect of a line failure can be calculated from the resupply switching sequence as shown in Figure \ref{resupply}. By assuming time constants for fault location, on-site switching and remote switching, we calculate the outage time for each of the $m$ stations in the case of a failure of each of the $n$ lines in the grid. By weighting the calculated outage time with the yearly failure rate, we get the expected yearly outage time $t_{out, i} [h/a]$ for a station $i$:
\begin{align}
 t_{out, i} = \sum_{k=0}^n h_{k} \cdot t_{out, ki} \label{eq:lineout}
\end{align}
where $h_{k} [1/a]$ is the yearly failure rate of line $k$ and $t_{out, ik} [h]$ is the outage time for station $i$ after a failure of line $k$. The yearly outage energy  $E_{out, i}[kWh/a]$ for a station $i$ can then be calculated with the installed power $P_k$ as:
\begin{align}
 E_{out, i} = P_k \cdot t_{out, i}
\end{align}
The ASIDI of the whole grid is then defined as the sum of all the outage energy in all $m$ stations in relation to the total installed power:
\begin{align}
 ASIDI = \frac{\sum_{i=0}^m E_{out, i}}{\sum_{i=0}^m P_{i}}
\end{align}

To prevent any decline in service reliability, we define the following reliability constraints:

\paragraph*{System wide Criterion} To assure that the target grid topologies are at least as reliable as the current grid, the ASIDI of the grid must not increase compared to the current grid.

\paragraph*{Station Criterion} To assure that no individual station suffers a great setback in service reliability compared to the current grid, the expected outage energy for each station is limited to an allowed maximum of $E_{out, max}$. For stations that already violate this constraint in current grid configuration, no further increase of the outage energy is allowed. In this case study, we use a maximum outage energy of $E_{out, max}=\unit[150]{kWh/a}$.

\subsubsection{Automating Resupply}
Secondary substations can be equipped with a communication link to allow remote controllable switching of the sectioning point. This measure accelerates the fault isolation and resupply process, so that it can be used to improve the service reliability. If there are violations of the reliability constraints in a feeder, the secondary substation in the load centre of the feeder is chosen for automation. This results in a partitioning, which confines the spread of the fault area and speeds up the resupply process. secondary substations are repeatedly selected until all constraints are met.
The resulting grids now comply with all planning constraints and are used as the reference grid for the radial and switching station grid respectively.

\subsection{Phase 4: Meshing} \label{sec:phase4}
In a last step, we construct a closed ring structure from the radial grid. Closed rings are only allowed with a more sophisticated protection system as detailed in Section \ref{sec:concepts}. The costs for updating the protection system are considered as per Table \ref{tab:costs}. In return, the meshing constraint is now relaxed so that the meshing of two feeders is allowed even without a switching station. The Iterated Local Search meta-heuristic is used to create a closed ring solution from each of the fifty valid radial topologies. As explained in Section \ref{sec:concepts}, closing sectioning points leads to a decrease in service reliability, so that compliance with the reliability constraints cannot be guaranteed after sectioning points are closed. We therefore demand that for every sectioning point that is closed, the associated secondary substations has to be automated. Since the open sectioning point is replaced with an automated sectioning point, the grid is considered at least as reliable as the radial grid.

The optimization creates fifty closed ring grids for each grid area and the cost minimal grid is chosen as the reference grid for the closed ring concept.

\subsection{Phase 5: Cost Comparison} 
We have constructed fifty grids for each of the three meshing concepts which all fulfil the planning constraints, but differ in their structure and therefore in their total costs. Figure \ref{cost_comparison}, shows the costs of the fifty topologies for each grid concept on the left. We select the cost minimal grid for each concept as the reference grid. These three grids are then compared to draw a conclusion about the differences of the three concepts. This comparison is shown for the example area in Figure \ref{cost_comparison} on the right. We can see, that the best open ring grid is cheaper than the best switching station solution for the example grid by around \unit[31,400]{\euro} per year. We can also see, that the best closed ring solution is more expensive than the best open ring solution, since the additional cost for secondary equipment is greater than the cost savings in grid reinforcement. We therefore conclude, that the best concept for the example grid area is a radial grid without a switching station.

\section{Results} \label{sec:results}
The methodology outlined in Section \ref{sec:optimization} for one grid area is applied to all switching stations in the case study grid. Of the 49 switching stations, 5 where dismissed due to faulty or inconsistent data. The methodology described in Section \ref{sec:optimization} is applied to all remaining 44 switching station grid areas. As 50 topologies are calculated for all the three network concepts in each of the 44 grid areas, a total of $3\cdot 50 \cdot 44 = 6600$ independent network plans are compiled for this study. To provide enough computational power, calculations were carried out on a computation cluster and over 840 million power flows were conducted to obtain the results. The results are now interpreted to draw a conclusion about the overall cost efficiency of the different MV grid concepts.

\subsection{Switching Stations}
We compare radial and switching station grids to evaluate the profitability of switching stations. The switching station improves reliability because of the included protection systems as well as the operational behaviour by allowing closed line rings. On the other hand, the costs for the switching station are equivalent to over \unit[5]{km} of cable \begin{figure}[h]
    \centering
		\includegraphics[width=1.\linewidth]{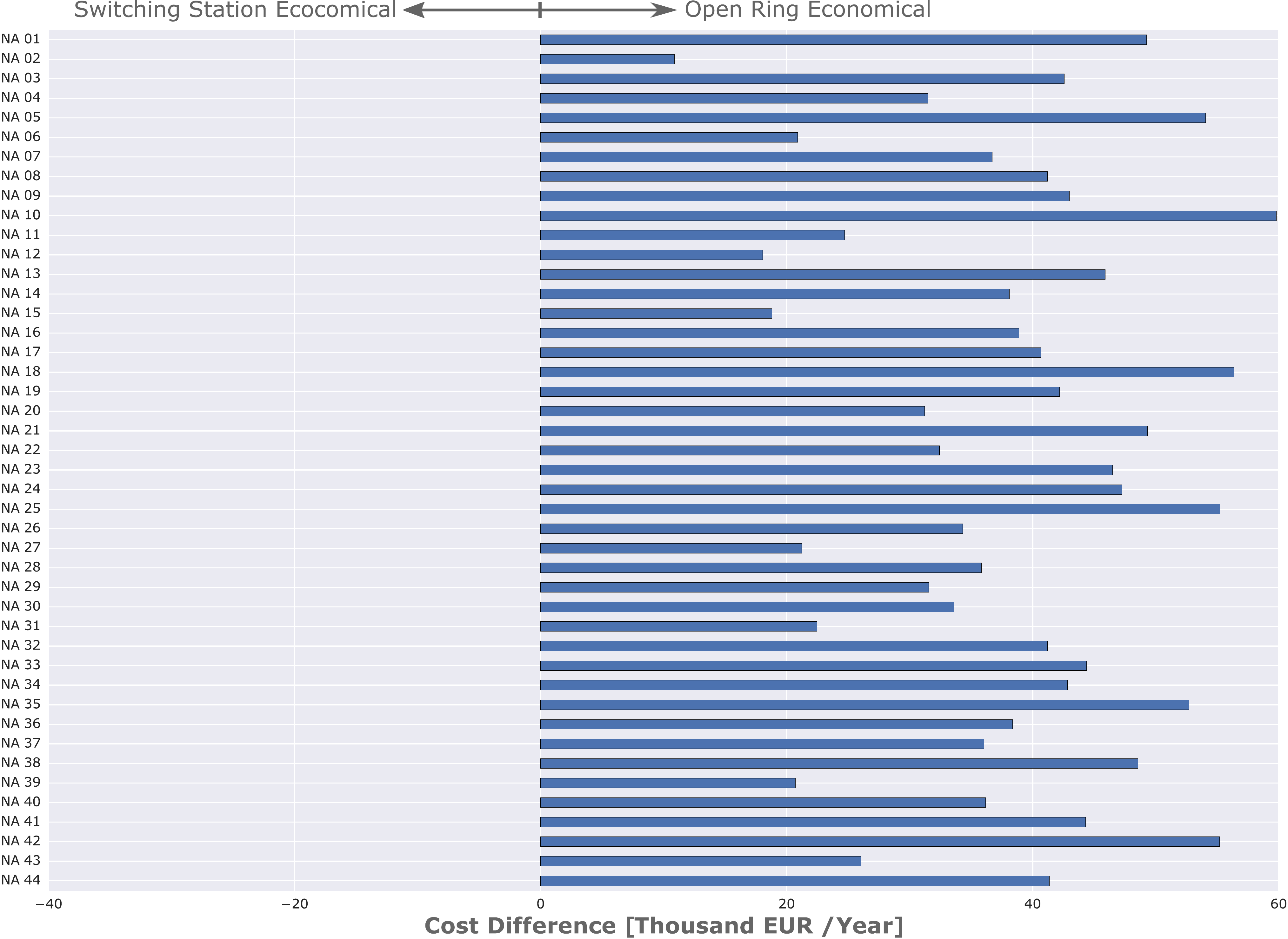}
    \caption{Cost difference between switching station and radial grid for all 44 grid areas (GA)}
    \label{result_swk}
\end{figure}
\begin{figure*}[h]
    \centering
		\includegraphics[width=.65\linewidth]{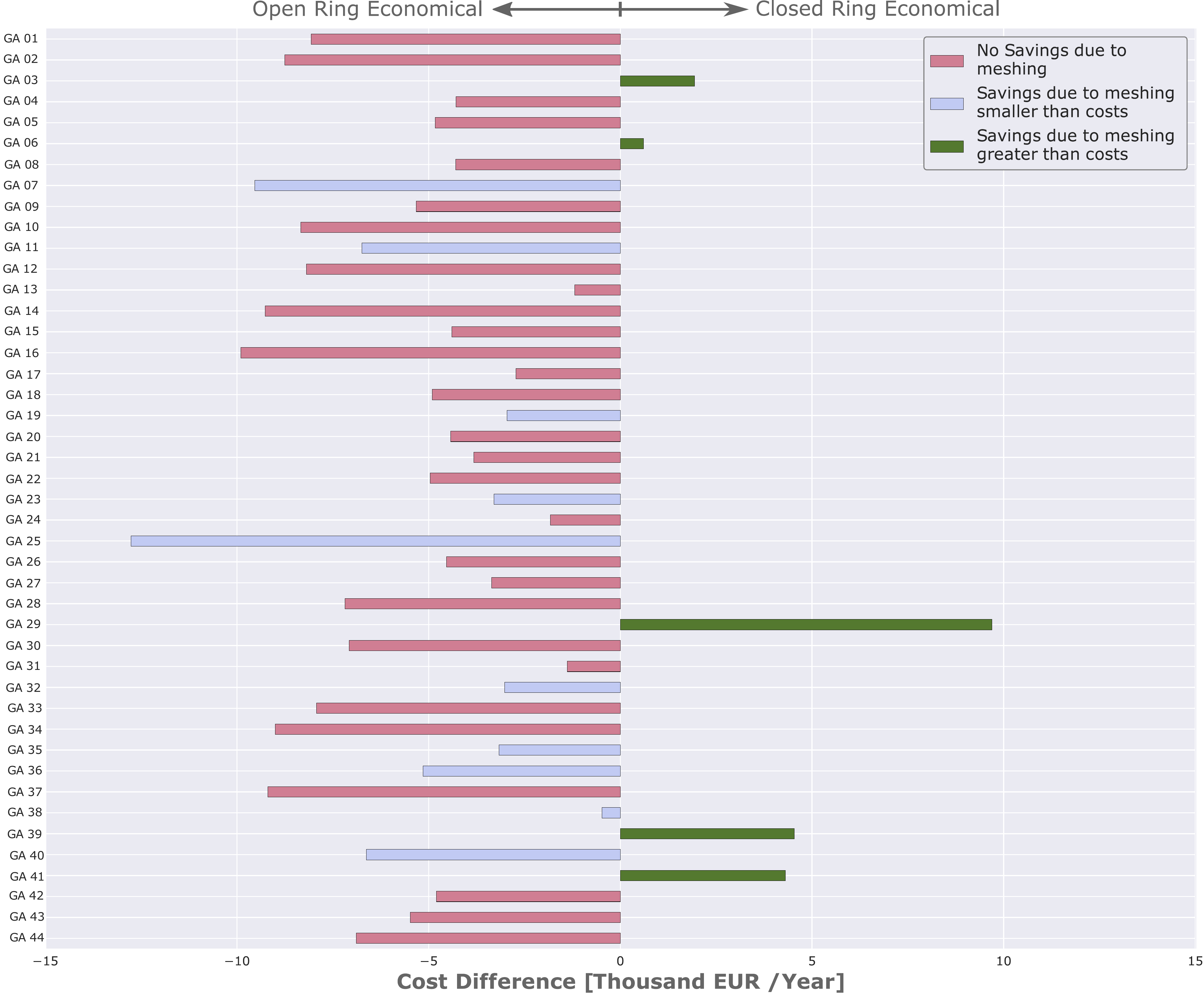}
    \caption{Cost difference between closed ring and radial grid for all 44 grid areas (GA)}
    \label{result_meshing}
\end{figure*}
and it needs additional supply lines to create a junction in the grid.
The comparison in Figure \ref{result_swk} shows, that the radial grid is more cost efficient in all 44 analysed cases.
This means that none of the switching stations are able to save enough grid reinforcement for it to be cost-efficient. Reliability only seems to be a minor impact factor, since compliance with the reliability constraints can be achieved with 2.4 automated stations on average, which is the cost equivalent of \unit[0.4]{km} of cable.

We therefore conclude, that a radial grid with automated secondary substations is preferable to a structure with switching stations. The radial grid structure with secondary substation automation is equivalent to the switching station grid in relation to grid operation and service reliability, but comes with significantly lower investment and maintenance costs.

\subsection{Closed Rings}
We now compare the closed ring grids with the open ring grids to evaluate the profitability of a closed ring concept. The comparison for all 44 cases is shown in Figure \ref{result_meshing}. In five grid areas, the closed ring solution is economical (green). In ten grids, the costs for the upgrade of the protection and fault location systems exceed the savings in grid reinforcement (orange). In the remaining 29 grids, closed rings do not lead to any savings in grid reinforcement and are therefore not cost-efficient (red). In conclusion, closed rings are only cost-efficient in very few grids, and even in those grids the saving potential with less than \unit[4,000]{\euro} per year in average is not very high. Since the DSO usually aims to have the same grid concept in all MV grids to allow harmonized procedures in grid planning and operation, the radial structure is more cost effective than the closed ring structure for the overall grid.

\section{Conclusion} \label{sec:conclusions}
In this study we economically compare target grids for radial, closed ring and switching station grids to come to a conclusion about the optimal grid concept in 44 real MV grid areas for a planning horizon in 2030. The results show, that switching stations are economically inefficient under the assumed constraints, since the technical benefits are outweighed by the costs for the switching station itself and the necessary supply lines. We therefore conclude that it is more cost-efficient to develop the grid towards a radial structure in the long run, avoiding replacement investments for switching stations that have reached the end of their life cycle. The closed ring concept is only cost effective in few grid areas, where the expected savings are small compared to the additional cost in other grid areas. We therefore conclude, that the radial grid in combination with selective automation of secondary substations is the most cost efficient mode of operation for the analysed grids.
The contributions of this paper can be summarized as follows: 
\begin{enumerate}
\item \textit{Automated target grid planning:} the methodology presented in this paper allows to automatically compile target grid plans based on geographical information and a wide range of technical constraints. This includes constraints for grid topology, normal operation power flow parameters, contingency operation power flow parameters and service reliability. 
\item \textit{Proof of concept:} the successful application of the approach to 44 grid areas with about \unit[4800]{km} of lines demonstrates the practical applicability of the approach. The high degree of automation makes the approach feasible for large-scale grid studies as well as for grid planning assistance systems.
\item \textit{Comparison of grid concepts:} The results of the case study allow to draw conclusions about the techno-economical differences of different MV grid concepts. A higher degree of meshing for the integration of DG was not found to be cost-efficient, as the additional costs for secondary equipment outweighed the cost savings in network reinforcement. 
\end{enumerate}
It should be noted that the conclusion about the grid concepts are only valid for the considered grids and under the considered boundary conditions. Even though the grids where chosen with the aim to include structurally different grids, a generalization for other grid groups is not easily possible. The boundary conditions with regard to protection system layout, spatial distribution of loads and DG or ratio of underground cables and overhead lines might differ significantly in other grids, especially of a different DSO. While meshed operation for the integration of DG could still be cost-efficient in other grids, our study does not confirm the assumption that meshing is always better for integration of DG. Instead, a comprehensive consideration of expected grid extension cost and protection system layout is necessary to determine the best mode of operation for the future power system. The approach could be further extended in the future to include transformation paths from the current power system to a future target grid or probabilistic scenarios for DG installation.

\section*{Acknowledgement}
This work was supported by the German Federal Ministry for Economic Affairs and Energy and the Projekttr\"ager J\"ulich GmbH (PTJ) within the framework of the project \textit{Smart Grid Models} (FKZ: 0325616).

\section*{References}
\bibliographystyle{elsarticle-num}
\bibliography{bibtex}
\end{document}